\documentclass[twocolumn,epjb,superscriptaddress]{revtex4}

\usepackage{dcolumn}
\usepackage{amsmath}

\usepackage{graphicx}

\begin{document}

\newlength{\figurewidth}
\setlength{\figurewidth}{0.95\columnwidth}
\setlength{\parskip}{0pt}
\setlength{\tabcolsep}{6pt}
\setlength{\arraycolsep}{2pt}

\title{Preservation of Network Degree Distributions From Non-Uniform Failures}
\author{Brian Karrer}
\affiliation{Department of Physics, University of Michigan, Ann Arbor, MI,
48109}
\author{Gourab Ghoshal}
\affiliation{Department of Physics, University of Michigan, Ann Arbor, MI, 48109}
\affiliation{Michigan Center for Theoretical Physics, University of Michigan, Ann Arbor, MI, 48109}

\begin{abstract}
There has been a considerable amount of interest in recent years on the robustness of networks to failures. Many previous studies have concentrated on the effects of node and edge removals on the connectivity structure of a \emph{static} network; the networks are considered to be static in the sense that no compensatory measures are allowed for recovery of the original structure. Real world networks such as the world wide web, however, are not static and experience a considerable amount of turnover, where nodes and edges are both added and deleted.  Considering degree-based node removals, we examine the possibility of preserving networks from these types of disruptions.  We recover the original degree distribution by allowing the network to react to the attack by introducing new nodes and attaching their edges via specially tailored schemes.  We focus particularly on the case of non-uniform failures, a subject that has received little attention in the context of evolving networks. Using a combination of analytical techniques and numerical simulations, we demonstrate how to preserve the \emph{exact} degree distribution of the studied networks from various forms of attack.
\end{abstract}

\pacs{89.75.Fb, 89.75.Hc}

\maketitle

Recent years have witnessed a substantial amount of interest within the physics community in the properties of networks ~\cite{AB02,DM02,Newman03d}.  Techniques from statistical physics coupled with the widespread availability of computing resources have facilitated studies ranging from large scale empirical analysis of the worldwide web, social networks, biological systems, to the development of theoretical models and tools to explore the various properties of these systems~\cite{AB01,SW01,WM01}. 

A relatively large body of work has been devoted to the study of degree distributions of networks, focusing both on their measurement, and formulation of theories to explain their emergence and their effects on various properties such as resilience and percolation.  These studies are mostly aimed at networks in the real world that evolve naturally, in the sense that they are driven by dynamical processes not under our control.  Representative examples being social, biological networks and information networks like the world wide web, which though manmade, grows in a distributed fashion.  There are however different classes of infrastructure related networks such as the transportation and power grids, communication networks such as the telephone and internet, that evolve under the direction of a centrally controlled authority.  

In addition to these is a relatively new class of networks which fall in between these two types, the classic example being peer-to-peer file-sharing networks.  These networks grow in a collaborative, distributed fashion, so that we have no direct influence over their structure.  However, we can manipulate some of the rules by which these form, giving us a limited but potentially useful influence over their properties.  It is a well established fact, that the structure of such networks is directly related to their performance. In view of this, a certain degree of effort has been made to tailor these \emph{designer} networks towards structures that optimize certain properties such as robustness to removal of nodes and efficient information transfer among other things~\cite{Hav-stan01,Ghoshal_Newman01}.  These networks typically experience a significant amount of vertex/edge turnover,  with users joining and leaving the network voluntarily, possible failures of key components and resources, or intentional attacks such as Denial of Service.  These factors can lead to severe disruption of the network structure and as a result, loss of its key properties.  In the face of this, it is natural to extend our analysis to the effects of these failures/attacks and use our limited control to attempt to adaptively restore the original structure of these networks.

Previous work has  focused on the effects of disruption on static networks, where authors have studied the connectivity structure under the random/targeted removal of nodes and edges~\cite{Cohen_EDH_2000,Broder01,Callaway_Newman01}.  The network is considered static in that no compensatory measures, such as the introduction of new edges or nodes, are permitted.  The effect of these removals have been measured against the existence of the \emph{giant component}: the largest set of vertices in the network of O($n$), where $n$ is the number of nodes, that are connected to each other by at least one path.  A representative example can be found in the paper by Albert \emph{et al}~\cite{Albert_Jeong01}, where they studied the size of the giant component of scale free networks such as the internet, under simulated random failures and targeted attacks on high degree nodes. One of the interesting things they found was that, while these networks were remarkably robust to random failures, they were extremely fragile to targeted attacks. This emphasizes the importance of non-uniform removal strategies. 

Unlike in the static case, the networks considered in this paper evolve in time with sustained node and edge removals.  The network is allowed to react to these disruptions via the introduction of new nodes and edges, chosen to be attached in a manner such that the network retains it original form, at least in terms of the degree distribution.  Such models, conventionally referred to in the literature as \emph{reactive networks} have been discussed before, see~\cite{Motter01,Roy02} for instance.  Here we assume that the designers of the network are only aware of the statistical properties of the removed nodes and have no ability to influence the existing network beyond the introduction of new nodes or reattachment of those removed.  Consequently they have two processes under their control to compensate for the attack.  The first is the degree of the introduced vertices and the second is the process by which a newly introduced vertex chooses to attach to a previously extant vertex on the network.  Failure is thus compensated by adding nodes and edges chosen from an appropriate degree distribution and attaching them to the network via specially tailored schemes.  Note that in our model, one can re-introduce nodes that have been removed or introduce completely new sets of nodes.  The former case could be indicative of say a computer in a peer-to-peer network that loses its connection, and would like to reconnect. The latter could represent the permanent loss of web-pages from the world wide web and the introduction of a new web-page.  We use the attachment kernel of Krapivsky and Redner~\cite{Red_Krap01}, to simulate the introduction of nodes and edges, and via the introduction of a deletion kernel we analyze the interesting and neglected case of non-uniform deletion. 

A variety of models have been proposed to simulate network evolution and growth where vertices are both added and deleted~\cite{Sarshar_Roy01,Chung_Lu01,Bennaim01,MGN01,Saldana01}, but these have concentrated on the relatively simple case of uniform deletion.  We will show that under uniform failures, the appearance of degree-degree correlations, that typically arise as a result of growth processes, as discussed in~\cite{DM02}, can be neglected. Previous models have taken advantage of precisely this fact to circumvent the difficulty of dealing with degree-degree correlations. For the case of non-uniform deletion, correlations cannot be ignored. In this paper we confront this issue by demonstrating how to preserve an initially uncorrelated network throughout the evolution process with the introduction of an additional rate equation for the degree-degree correlations.  We give analytical results and numerical simulations for a variety of degree distributions under various forms of attack. In all the cases that we study, we recover the \emph{exact} degree distributions.
 
\section{The Model}
\label{sec:rate}   
 
Consider a network which evolves under the removal and addition of vertices.  In each unit of time 
we add $1$ vertex and remove $r$ vertices.  Removal of a vertex also implies that all the edges incident on that vertex vanish and consequently the degree of vertices at the end of those edges decrease.  Here $r$ can be interpreted as the ratio of vertices removed to those added, so $r < 1$ represents a growing network, $r > 1$ a shrinking one, while $r = 1$ implies vertex turnover but fixed network size.  The equations to follow represent the completely general case. However, for the purposes of this paper we will specialize to networks of constant size as we assume that the network already exists and we would like to preserve its original structure, by balancing the rate of attack against the rate of repair.

Let $p_k$ be the fraction of nodes in the network that at a given time have degree $k$. By definition then it has the normalization:

\begin{equation}
\sum_{k} p_k= 1.
\end{equation}
In addition to this we would like to have freedom over the degree of the incoming vertex.  Let $m_k$ be the probability distribution governing this, with the constraint $\sum_k k m_k = c$.  We also have to consider how a newly arriving vertex chooses to attach to other vertices extant in the network and how a vertex is removed from the same.  Let $\pi_{k}$ be the probability that a given edge from a new node is connected to a node of degree $k$, multiplied by the total number of nodes $n$.  Then $\pi_{k}p_k$ is the probability that an edge from a new node is connected to some node of degree $k$.  Similarly, let $a_k$ be the probability that a given node with degree $k$ fails or is attacked during one node removal also multiplied by $n$.  Then $a_k p_k$ is the total probability to remove a node with degree $k$ during one node removal.  Note that the introduction of the deletion kernel $a_k$ is what sets our model apart from previous models describing the network evolution process.  Since each newly attached edge goes to some vertex with degree $k$, we have the following normalization conditions:
\begin{eqnarray}
&\sum_{k}& \pi_{k} p_k = 1, \\
&\sum_{k}& a_k p_k = 1. 
\end{eqnarray}

\subsection{Rate Equation}

Armed with the given definitions and building on the work done previously by~\cite{MGN01}, we are now in a position to write down a rate equation governing the evolution of the degree distribution.  For a network of $n$ nodes at a given unit of time, the total number of nodes with degree $k$ is $np_k$.  After one unit of time we add one vertex and take away $r$ vertices, so the number is $(n + 1 - r)p_k'$, where $p_k'$ is the new value of $p_k$.  Therefore we have,
\begin{eqnarray}
(n+1-r)p_k' &=& n p_k
	\nonumber\\
	& & {}  + c\pi_{k-1} p_{k-1}
	\nonumber\\
	 & & {} - c\pi_{k}p_k
	\nonumber\\
	& & {} + r\sum_{j} e_{k+1|j}j a_j p_j 
	\nonumber\\
	& &{} - r\sum_{j} e_{k|j} j a_{j} p_{j} 
	\nonumber\\
	& &{} - ra_k p_k+m_k,
\label{eq:rate}	
\end{eqnarray}
where $e_{k|j}$ is the conditional probability of following an edge from a node of degree $j$ and reaching a node of degree $k$.  Alternatively, it is the degree distribution of nodes at the end of an edge emanating from a node of degree $j$.  Note that $e_{0|j}$ and $e_{j|0}$ are always zero, and for an uncorrelated network, $e_{k|j}= k p_k / \langle k \rangle$.  The terms involving $\pi_{k}$ describe the flow of vertices with degree $k-1$ to $k$ and $k$ to $k + 1$ as a consequence of edges gained due to the addition of new vertices.  The first two terms involving $a_j$ describes the flow of vertices with degree $k+1$ to $k$ and $k$ to $k-1$ as vertices lose edges as a result of losing neighbors.  The term $-r a_{k} p_{k}$ represents the direct removal of a node of degree $k$ at rate $r$.  Finally $m_k$ represents the addition of a vertex with degree $k$.  Processes where vertices gain or lose two or more edges vanish in the limit of large $n$ and are not included in Eq.~\eqref{eq:rate}.

The rate equation described above presents a formidable challenge due to the appearance of $e_{k|j}$ from the terms representing deleted edges from lost neighbors.  Rate equations for  recovery schemes based on edge rewiring are slightly easier to deal with.  Upon failure, all edges connected to that node are rewired so that the degrees of the deleted node's neighbors do not change, and this term does not appear.  The specific case of preferential failure in power-law networks was considered previously in this context by~\cite{Roy02}.   However, this recovery protocol can only be used on strictly growing networks, because a network of constant size would become dense under its application. Moreover, it is dependent on the power-law structure of the network.  The methods described here are general and are applicable to arbitrary degree distributions.

Apart from edge rewiring, the special case of random deletion also leads to a significant simplification.  Uniform deletion amounts to setting $a_k = 1$.  Doing so, then leads to the following,
\begin{eqnarray}
\sum_{j} e_{k|j}j p_{j}=kp_{k}, 
\label{eq:random}
\end{eqnarray}
which renders Eq.~\eqref{eq:rate} independent of $e_{k|l}$ and thus independent of any degree-degree correlations.  Random deletion hence closes equation~\eqref{eq:rate} for $p_{k}$, enabling us to seek a solution for the degree distribution for a given $m_{k}$ and $\pi_{k}$.  With non-uniform deletion, the degree distribution depends on a two-point probability distribution, and as we shall see in Section~\ref{sec:correlations}, the two-point probability distribution will depend on the three-point probability distribution and so on.  This hierarchy of distributions, where the $n$-point distribution depends on the $n+1$-point distribution, is not closed under non-uniform failure and hence it is difficult to seek an exact solution for the degree distribution.  Nevertheless, in the following, we demonstrate a method that allows us to navigate our way around this problem.
    
As mentioned before, for the purposes of this paper we will be interested in a network of constant size, where the rate of attack is compensated by the rate of repair.  Assuming that the network reaches (or already is) a stationary distribution and does not possess degree-degree correlations, we set $r = 1$ and can further simplify Eq.~\eqref{eq:rate}. Let $\left\langle k \right\rangle_{a}$ be the mean degree of nodes removed from the network (i.e. $\left\langle k \right\rangle_{a}=\sum_{k} ka_{k}p_{k}$), and $\left\langle k \right\rangle$ the mean degree of the original degree distribution $p_{k}$.  Then we have,
\begin{eqnarray}
0 &=& c\pi_{k-1}p_{k-1} \nonumber\\ &-&c\pi_{k}p_{k}+(k+1)\frac{\left\langle k \right\rangle_{a}}{\left\langle k \right\rangle}p_{k+1} \nonumber\\&-&k\frac{\left\langle k \right\rangle_{a}}{\left\langle k \right\rangle}p_{k}-a_{k}p_{k}+m_{k}.
\label{eq:rate3}
\end{eqnarray}
The evolution process, specifically non-uniform removal of nodes, can and in many cases will introduce degree-degree correlations into our networks.  In order to confront this issue, we will first find choices for $m_k$ and $\pi_k$ that satisfy the solutions to the rate equation, for a given $p_k$, in a network that is uncorrelated.  We will then demonstrate that a special subset of those solutions for $m_{k}$ and $\pi_{k}$ is an uncorrelated fixed point of the rate equation for the degree-degree correlations.  This opens up the possibility, that a network that initially has no degree-degree correlations will not develop correlations from the evolution process.

Although the rate equation described in Eq.~\eqref{eq:rate3} is fairly complicated, it is a relatively straightforward exercise to determine the relation between edges added to those removed.  Multiplying Eq.~\eqref{eq:rate3} by $k$, summing over $k$ and rearranging yields $\left\langle k \right\rangle_{a} = c$.  This equation is simple to interpret.  Since the network has a constant fixed-point degree distribution, the average degree of the network remains constant, and therefore edges are removed and added at the the same rate.  

\section{Recovery from attacks}

In this section we describe our method under which networks can recover from various forms of attack.  The types of attack we consider are those studied generally by most authors (though in static networks), namely preferential and targeted attacks.  

Random failures are the most generally studied schemes in both static and evolving networks, in view of the fact that they lend themselves to relatively simple analysis.  These types of failures may be representative, say, of disruption of power lines or transformers in a power grid owing to extraneous factors such as weather.  However, the functionality of most networks often depends on the performance of higher degree nodes, consequently non-uniform attack schemes focus on these.  For example, in a peer-to-peer network, a high degree node could be a central user with large amounts of data.  High degree could also be indicative of the amount of load on a node during its operation, or on the public visibility of a person in a social network.  It is reasonable to assume that a malicious entity such as a computer virus is more likely to strike these important nodes.  Holme \emph{et al}~\cite{Holme_Kim01} have employed this removal strategy (among others) on a variety of simulated and real networks and have found it to be highly effective in disrupting the structure of the attacked network.

\begin{figure}[h]
\includegraphics[width=8cm]{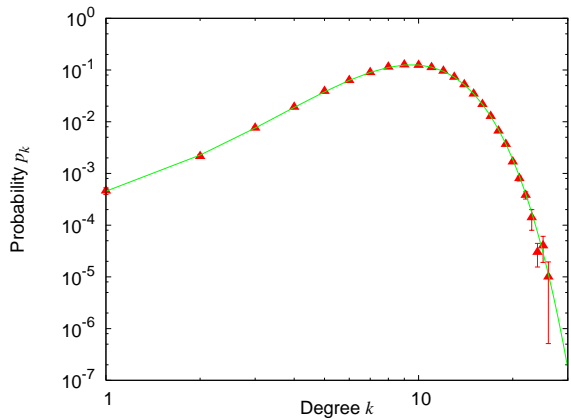}
\caption{Degree distribution of a Poisson network ($10^{4}$ nodes) with mean $\mu = 10$, under preferential attack $a_{k} \propto k$ and uniform attachment $\pi_{k}=1$ using $m_{k}=a_{k}p_{k}$.}
\label{fig:poissonpref}
\end{figure}

We simulate these kinds of attacks using preferential failure $a_{k} \propto k$, that sample nodes in proportion to their number of connections, and through an outright attack on the highest degree nodes represented by $a_{k} \propto \theta(k-k_{min})$, where $\theta(x)$ is the Heaviside step function. 
Our method of compensation will involve control over two processes: the first where our newly incoming/repaired vertex chooses a degree for itself drawn from some distribution $m_k$, and second, the process by which this vertex decides to attach to any other vertex in the network, governed by the attachment kernel $\pi_k$.
 
\subsection{Using $m_k$ and the attachment kernel $\pi_{k}$}
\label{sec:designattachment}

Our goal here is to solve for the attachment kernel $\pi_{k}$, that will preserve the original probability distribution $p_{k}$, subject to a deletion kernel $a_k$ for some choice of $m_{k}$.  We will assume that the final network is uncorrelated and work with Eq.~\eqref{eq:rate3}, keeping in mind that any arbitrary choice of $m_{k}$ and $\pi_{k}$ is probably not consistent with that assumption.  

Introducing the cumulative distribution for the attacked and newly added vertices, $A_{k}$ and $M_{k}$ respectively,
\begin{equation}
A_{k}  =  \sum_{l=k}^{\infty}a_{l}p_{l}, \qquad M_{k} = \sum_{l=k}^{\infty}m_{l}, 
\label{eq:cums}
\end{equation}
we sum Eq.~\eqref{eq:rate3} from $k = k'+1$ to $\infty$, noting that $\left\langle k \right\rangle_{a} = c$ for our steady state network. This leads to the following relation,
\begin{equation}
\pi_{k}p_{k} = \frac{(k+1)p_{k+1}}{\left\langle k \right\rangle} +\frac{A_{k+1}-M_{k+1}}{c}.
\end{equation}
Dividing both sides by $p_{k}$ gives us an expression for the attachment kernel, 
\begin{eqnarray}
\pi_{k} &=& \frac{1}{p_{k}} \left[\frac{(k+1)p_{k+1}}{\left\langle k \right\rangle}
+ \frac{A_{k+1} - M_{k+1}}{c}\right].
\nonumber\\
\label{eq:genattachment}
\end{eqnarray}

\begin{figure}
\includegraphics[width=8cm]{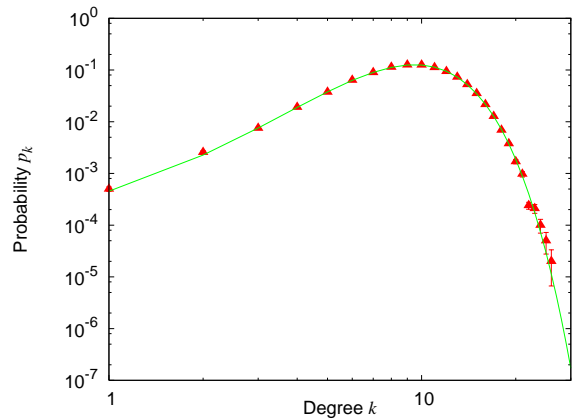}
\caption{Degree distribution of a Poisson network ($10^{4}$ nodes) with $\mu = 10$,  under high degree attack $a_{k} \propto \Theta(k-12)$ and uniform attachment $\pi_{k}=1$ using $m_{k}=a_{k}p_{k}$.}
\label{fig:poissontheta}
\end{figure} 

Equation~\eqref{eq:genattachment} represents the set of possible solutions for the attachment kernel that will lead to the desired degree distribution,  given that the final network is uncorrelated.  The correct choice of solution from the above set, must obey the consistency condition, that when inserted into the rate equation for the degree-degree correlations, the correlations vanish.  In Section~\ref{sec:correlations}, we will show that the following \textit{ansatz} chosen from the above set is such a choice:
\begin{eqnarray}
m_{k} &=& a_{k}p_{k}, \nonumber\\
\pi_{k}&=&\frac{(k+1)p_{k+1}}{\left\langle k \right\rangle p_k}.
\label{eq:attachment}
\end{eqnarray}
Equation~\eqref{eq:attachment} was previously derived by~\cite{Ghoshal_Newman01} for the case of  random deletion.  Here we posit that it works more generally for the case of non-uniform attack when our initial network is uncorrelated (with some caveats that will be explained shortly). 

The choice of $\pi_{k}$ makes intuitive sense because the quantity $(k+1)p_{k+1}/\left\langle k \right\rangle$ is the probability distribution governing the number of edges belonging to a node, reached by following a randomly chosen edge to one of its ends, \emph{not including} the edge that was followed.  This is one less than the total degree of the node and is also referred to as the \emph{excess} degree distribution.  Note that in our model we specify the degree of incoming nodes. Therefore the appearance of the excess degree distribution is a signature of an uncorrelated network, implying the newly arriving edges are being introduced in an uncorrelated fashion. 

There are basically two conditions for the existence of a solution given by Eq.~\eqref{eq:attachment}; $a_{k}p_{k}$ must be a valid probability distribution, and $\left\langle k \right\rangle$  must be finite.  These are not very stringent conditions and are typically satisfied by most degree distributions.  In other words, barring some pathological cases, it is always possible to find a solution of the form of Eq.~\eqref{eq:attachment}.  There is an additional consideration,  the deletion process may lead to nodes of degree zero in a network that originally did not have any such nodes.  While the fraction of such nodes is vanishingly small for networks with say, Poisson degree distributions, they may be non-trivial for power-law networks.  As such, it is important to set $\pi_{0}$ (the probability to attach to a node of degree zero) to a generous value in order to reconnect these nodes to the network.

We are now in a position to effect our repair on the network.  Given the original degree distribution $p_k$ and the form of the attack $a_k$, Eq.~\eqref{eq:attachment} gives us the precise recipe for recovering the degree distribution.  We need to sample the degrees of the newly introduced nodes in proportion to the product of the deletion kernel \emph{and} the degree distribution, and then attach these edges in proportion to the excess degree distribution of the network.  To test our repair method, we provide four examples for initially uncorrelated networks with $10,000$ nodes generated using the \emph{configuration model}~\cite{Molloy_Reed01,Luczak}.  In the configuration model, only the degrees of vertices are specified, apart from this sole constraint the connections between vertices are made at random.  
 
\begin{figure}
\includegraphics[width=8cm]{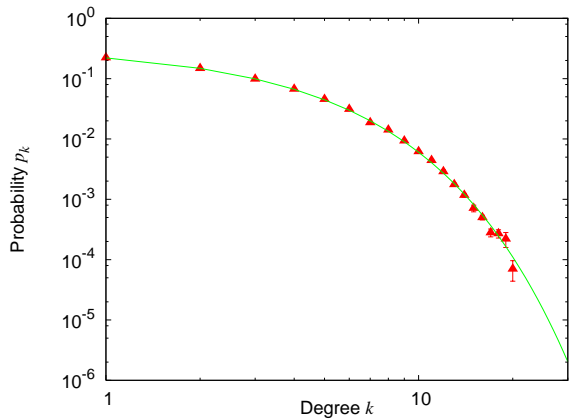}
 \caption{Degree distribution of an exponential network ($10^4$ nodes) with $\lambda = 0.4$ under targeted attack $a_{k} \propto \Theta(k-5)$ using  $\pi_{k}$ from Eq.~\eqref{eq:attachment} after setting $m_{k}=a_{k}p_{k}$.}
\label{fig:exptheta}
\end{figure}

The simulation results show the initial degree distribution and the compensated one subject to two types of attacks on Poissonian networks with degree distribution given by,
 \begin{equation}
 p_k = \frac{e^{-\mu} \mu^{-k}}{k!}.
 \label{eq:poisson1}
 \end{equation} 

In Fig. \ref{fig:poissonpref} we show the resulting degree distribution where nodes were attacked preferentially, i.e. $a_{k} \propto k$, while in Fig. \ref{fig:poissontheta} we show the case for targeted attack only on high degree nodes represented by $a_{k} \propto \Theta(k-k_{min})$ where $k_{min}$ is the \emph{minimum} degree of the node attacked. The degrees of newly added nodes were chosen from the distribution $a_{k}p_{k}$ with the attachment kernel $\pi_{k}$ set to one, corresponding to the solution of equation~\eqref{eq:attachment} after substituting in the appropriate $p_k$.  The data points in all the figures are averaged over multiple realizations of the network each subject to $10^5$ iterations of addition and deletion. The points along with corresponding error bars represent the final degree distribution, whereas the solid line represents the initial network. As the figures show, the final networks are in excellent agreement with the initial degree distribution. 

\begin{figure}
\includegraphics[width=8cm]{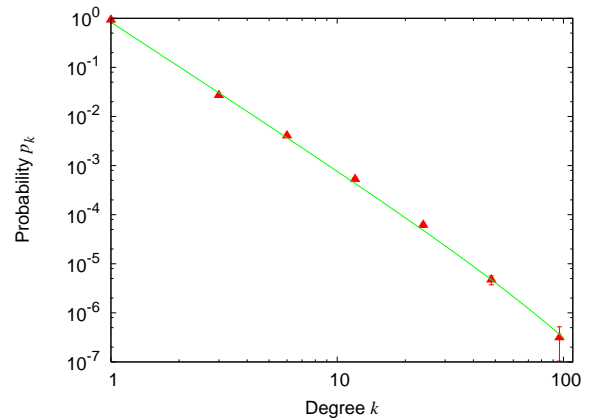}
  \caption{Log-binned degree distribution of a power law network ($10^4$ nodes) with exponent $\gamma = 3$ and exponential cutoff $\kappa = 100$, under preferential attack $a_{k} \propto k$ using $\pi_{k}$ from Eq.~\eqref{eq:attachment} after setting $m_{k}=a_{k}p_{k}$.}
\label{fig:powerpref}
\end{figure}

We employ the same attack kernels, $a_{k} \propto k$ and a targeted attack only on high degree nodes represented by $a_{k} \propto \Theta(k-k_{min})$ on two other examples. Our first example network has links distributed according to a power-law with an exponential cutoff,

\begin{equation}
p_{k} = \left\lbrace\begin{array}{ll}
           Ck^{-\gamma}e^{-k/\kappa} & \qquad\mbox{$k \neq 0$,}\\
           0 & \qquad\mbox{$k=0$}\\
         \end{array}\right.
\label{eq:powerexp}         
\end{equation}
$C$ is a normalization constant which in this case is $1/Li_{\gamma}(e^{-1/\kappa})$, where the function $Li_{\nu}(z)$ is the poly-logarithm function defined as:
\begin{equation}
Li_{v}(z) = \sum_{k=1}^\infty \frac{z^k}{k^\nu}.
\end{equation} 
The exponential cut-off has been introduced for three reasons.  First, many real world networks appear to show this cutoff~\cite{Amaral_Scala01} and second, it renders the distribution normalizable for ranges of the exponent $\gamma \le 2$. Finally, for a pure power-law network it is in principle possible to assign a degree to a node that is greater than the system size. The exponential cutoff ensures that the probability for this to happen is vanishingly small. In the other examples that we consider, the functional form of the distribution already ensures this property.

The second network has an exponential distribution given by,
\begin{equation}
p_{k} = (1-e^{-\lambda})e^{-\lambda k}.
\end{equation}

Fig. \ref{fig:exptheta} shows the results for the exponentially distributed network ($\lambda = 0.4$) undergoing targeted attack.  In Fig. \ref{fig:powerpref} we show the resulting degree distribution for the power-law network ($\gamma = 3$ and $\kappa = 100$) where nodes were attacked preferentially. Both figures indicate the initial and final networks are in excellent agreement. 

At this point, aside from the technical details, it is worth reminding ourselves of the big picture.  We have demonstrated above that if a network with a certain degree structure is subjected to an attack that aims to destabilize that structure, one can recover the same, by manipulating the rules by which vertices are introduced to the network.  The rules that we employ in our repair method are dependent on the types of attacks that our networks are subject to.  In the following section we give a detailed justification of the employment of our method.

\subsection{Neglecting degree-degree correlations}
\label{sec:correlations}

In order for our results from the previous sections to be valid,  we must demonstrate that our initially uncorrelated networks remain uncorrelated under our repair scheme.  To accomplish this, we will define a rate equation for the degree-degree correlations and demonstrate that the uncorrelated network is a fixed point of this equation.  Our rate equation will describe the evolution of the expected number of edges in the network with ends of degree $k$ and $l$.  

Let the expected number of such edges in the network be,
\begin{equation}
m e_{l,k},
\label{eq:edgerate}
\end{equation}
where $m = n \langle k \rangle / 2$ , and $e_{l,k}$ is the probability that a randomly  selected edge has degree $k$ at one end and degree $l$ in the other.  The expected number of edges after one time step where we add $c$ and take away $\langle k \rangle_{a}$ edges is then,
\begin{equation}
[m + c - \langle k \rangle_{a}] e'_{l,k} = m e_{l,k} + \Delta,
\label{eq:edgerate1}
\end{equation}
where $\Delta$ represents all other edge addition and removal processes. 
 
We have already established that in the steady state case, $\langle k \rangle_{a} = c$ irrespective of the degree distribution, so our goal is equivalent to showing that $\Delta$ is equal to zero for an uncorrelated network generated/repaired with our special choices of $\pi_{k}$ and $m_{k}$.  As a result  $e'_{k,l} = e_{k,l}$, implying that the degree-degree correlations (if any) remain constant over time.  

We will assume that our network is locally tree-like, something which holds true for most random graphs. In addition we will only consider processes out to second nearest-neighbors of a node.  These assumptions allows us to avoid including terms in the rate equation representing removal of nodes with neighbors that are connected to each other.   Nevertheless, there are a large number of remaining processes that we will need to consider.  

To start things off, note that the rate equation is symmetric in the indices $l$ and $k$.  Any process that contributes to changing $k$ while holding $l$ constant also contributes to changing $l$ while holding $k$ constant. We can therefore consider contributions to $\Delta$ from $e_{k-1,l}$, $e_{k,l}$ and $e_{k+1,l}$ and add on the corresponding symmetric terms at the end.  The first process we need to take into account is a direct addition of a node of degree $l$.  This contributes two flows to the rate equation, $l \pi_{k-1}p_{k-1}m_{l}$ and $-l \pi_{k}p_{k}m_{l}$.  Similarly, the direct deletion of a node of degree $l$ contributes $-l e_{k|l}a_{l}p_{l}$ and $l e_{k+1|l}a_{l}p_{l}$. Next, we will have to take into account second nearest-neighbor processes.  We can be certain that these terms are of the same order by merely counting the number of unsummed probability distributions  that go into each process.  There will be two terms for the attachment process representing the situation where a new node of any degree attaches to a node of degree $k$ or $k-1$,  that was previously attached to a node of degree $l$.  These terms are $-cke_{l|k}\pi_{k}p_{k}$ and $c(k-1)e_{l|k-1}\pi_{k-1}p_{k-1}$. Similarly there are two removal processes, where a node of any degree that is removed from the network was previously attached to a node of degree $k$ or $k+1$ that has neighbor(s) of degree $l$.  Unfortunately these terms introduce three-point correlations into the rate equation.  Analogous to methods employed in similar hierarchy problems, we use a moment-closure approximation to represent these processes as a product of two two-point correlations in the following manner,
\begin{equation}
-\sum_{j} (k-1)e_{l|k}e_{k|j} j a_{j} p_{j}+\sum_{j}k e_{l|k+1}e_{k+1|j}ja_{j}p_{j}.
\end{equation}
 
Adding all of these terms together our final equation for $\Delta$ is,
\begin{eqnarray}
\Delta &=& l \pi_{k-1}p_{k-1}m_{l}  - l \pi_{k}p_{k}m_{l} + l e_{k+1|l}a_{l}p_{l} - l e_{k|l}a_{l}p_{l}  \nonumber\\
&+&c(k-1)e_{l|k-1}\pi_{k-1}p_{k-1} - c k e_{l|k}\pi_{k}p_{k}\nonumber\\
&+&\sum_{j}k e_{l|k+1}e_{k+1|j}ja_{j}p_{j} - \sum_{j} (k-1)e_{l|k}e_{k|j} j a_{j} p_{j},
\nonumber\\
\label{eq:delta}
\end{eqnarray}
in addition to terms where $l$ and $k$ are interchanged.  

After inserting the appropriate $\pi_{k}$ and $m_{k}$ from Eq.~\eqref{eq:attachment} along with the uncorrelated solution $e_{k|l}= kp_{k} / \langle k \rangle$, it can be shown that,
\begin{equation}
\Delta=0.
\end{equation}
According to Eq.~\eqref{eq:edgerate1}, there exist a set of solutions such that an initially uncorrelated network will not develop any degree-degree correlations as a consequence of the evolution process.  The attachment kernel that was employed in the network evolution process, described in Section~\ref{sec:designattachment}, was a subset of these solutions.  This allowed the repair method to be employed by maintaining negligible correlations in the network.
 
One must point out, that we have not explicitly demonstrated the stability of the uncorrelated solution to perturbations.  For example fluctuations in $e_{k,l}$ or in the number of edges may drive the network away from the uncorrelated steady-state.  An analytical approach to determine this, say using linear stability analysis is difficult, due to the numerous related probability distributions involved.  So instead we resort to a numerical approach. We measured the Pearson correlation coefficient between the degrees of nodes at both ends of an edge for all our model networks.  For the Poisson and exponential cases, the correlations remained negligible during the evolution process.  On the other hand, the power-law network developed non-trivial correlations.  We have not been able to determine whether the appearance of these correlations was due to finite-size effects, or instability in the uncorrelated solution, or to some other cause.  The results show that the agreement between the initial and final degree distributions is very good, and it seems that in this particular case, the correlations did not demonstrate a significant effect on the final state of the network.  
 
\section{Conclusion}
\label{sec:conc}
In this paper, we have shown how to preserve a network's degree distribution from various forms of attack or failures by allowing it to adapt via the simple manipulation of rules that govern the introduction of  nodes and edges.  We based our analysis on a rate equation describing the evolution of the network under arbitrary schemes of addition and deletion.  In addition to choosing the degree of incoming nodes, we allow ourselves to choose how nodes attach to the existing network. To deal with the special case of non-uniform deletion we have introduced a rate equation for the evolution of degree-degree correlations and have used that in combination with the equation for the degree distribution to come to our solution.  We have provided examples of the applicability of this method using a combination of analytical techniques and numerical simulations on a variety of degree distributions, yielding excellent results in each case. 

The structure of many networks in the real world is crucially related to their performance. Many authors have seized on the fact that technological networks such as the internet and peer-to-peer networks are power-law in nature, and have used this to design efficient search schemes among other things.  Loss of structural properties of these networks then lead to severe constraints on their performance. Recent empirical studies~\cite{Roy03} have suggested that node removal, for example, in the world wide web, is typically non-uniform in nature. In view of this, it is crucial for researchers to come up with effective solutions to try and manage these types of disruptions.  To the best of our knowledge, there is a considerable gap in understanding the non-uniform deletion process of nodes and edges and corresponding methods to deal with them. This paper begins to address this gap.
   
It must be pointed out that the methods we have described depends crucially on the assumption of  negligible correlations as the network evolves. Curiously enough, in our example power-law network, we were able to get very good agreement between the initial and final degree distributions, in spite of the appearance of non-trivial correlations. It will certainly be interesting to see if our methods can be extended to the case of networks with strong correlations, and other metrics describing network structure.  Perhaps it is possible to directly confront the rate equation for the degree-degree correlations, although this seems a difficult prospect at the moment.  The idea of preserving the structure of networks from attacks by allowing it to react in real-time is a relatively nascent one and the authors look forward to more developments in this area.

The authors thank Mark Newman for illuminating discussions.  This work was funded by the James S. McDonnell Foundation.

\bibliographystyle{apsrev}

\begin{thebibliography}
\expandafter\ifx\csname url\endcsname\relax
  \def\url#1{\texttt{#1}}\fi
\expandafter\ifx\csname urlprefix\endcsname\relax\def\urlprefix{URL }\fi

\bibitem{AB02}
R.~Albert and A.-L. Barab\'asi, Rev. Mod. Phys. \textbf{74}, 47 (2002).

\bibitem{DM02}
S.~N. Dorogovtsev and J.~F.~F. Mendes, Adv.
Phys. \textbf{51}, 1079 (2002).

\bibitem{Newman03d}
M.~E.~J. Newman, SIAM Review \textbf{45}, 167 (2003).

\bibitem{AB01}
A.~-L.~ Barab\'asi and R.~Albert, Science \textbf{286}, 509 (1999).

\bibitem{SW01}
D.~J.~Watts and S.~H.~Strogatz, Nature \textbf{393}, 440 (1998).

\bibitem{WM01}
R.~J. Williams and N.~D.~Martinez, Nature \textbf{404}, 180 (2000).

\bibitem{Hav-stan01}
G. ~Paul, T.~Tanizawa, S.~Havlin and H.~E. Stanley, Eur. Phys. J. B \textbf{38}, 187 (2004).

\bibitem{Ghoshal_Newman01}
G.~Ghoshal and M.~E.~J.~Newman, Eur. Phys. J. B, \textbf{58}, 175 (2007).

\bibitem{Cohen_EDH_2000}
R.~Cohen, K.~Erez, D.~ben-Avraham, and S.~Havlin, Phys. Rev. Lett. \textbf{85}, 4626 (2000).

\bibitem{Broder01}
A.~Broder, R.~Kumar, F.~Maghoul, P.~Raghavan, S.~Rajagopalan, R.~Stata, A.~Tomkins and J.~Wiener, Comput. Netw. \textbf{33}, 309 (2000).

\bibitem{Callaway_Newman01}
D.~S.~Callaway, M.~E.~J.~Newman, S.~H.~Strogatz, and D.~J.~Watts, Phys. Rev. Letts. \textbf{85}, 5468 (2000).

\bibitem{Albert_Jeong01}
R.~Albert, H.~Jeong and  A.-L. Barab\'asi, Nature \textbf{406}, 378 (2000).

\bibitem{Motter01}
A.~E.~Motter, Phys. Rev. Letts. \textbf{93}, 098701 (2004).

\bibitem{Roy02}
B.~Rezai, N.~Sarshar, V.~Roychowdhury, P.~Oscar Boykin, Physica A \textbf{381}, 497 (2007).

\bibitem{Red_Krap01}
P.~L.~Krapivsky and S.~Redner, Phys. Rev. E \textbf{63}, 066123 (2001).

\bibitem{Sarshar_Roy01}
N.~Sarshar and V.~Roychowdhury, Phys. Rev. E \textbf{69}, 026101 (2004).

\bibitem{Chung_Lu01}
F.~Chung and L.~Lu, Internet Mathematics \textbf{1}, 463 (2004).

\bibitem{Bennaim01}
E.~Ben-Naim and P.~L.~Krapivsky, J. Phys. A: Math. Theor. \textbf{40} 8607 (2007).

\bibitem{MGN01}
C.~ Moore, G.~Ghoshal, M.~E.~J.~ Newman, Phys. Rev. E \textbf{74}, 036121(2006).

\bibitem{Saldana01}
J.~ Salda\~{n}a, Phys. Rev. E \textbf{75}, 027102 (2007).

\bibitem{Holme_Kim01}
P.~Holme, B.~J.~Kim, C.~N.~Hoon and S.~K.~Han, Phys. Rev. E \textbf{65}, 056109 (2002).

\bibitem{Molloy_Reed01}
M.~Molloy and B.~Reed, Random Structures and Algorithms \textbf{6}, 161 (1995).

\bibitem{Luczak}
T.~Luczak in \emph{Proceedings of the Symposium on Random Graphs, Pozna\'n 1989}, edited by T.~M.~Frieze and T.~Luczak (John Wiley, New York 1992), pp.165-182.

\bibitem{Amaral_Scala01}
L.~A.~N.~Amaral, A.~Scala, M.~Barth\'el\'emy, and H.~E.~Stanley, Proc. Natl. Acad. Sci. \textbf{97}, 11149 (2000).

\bibitem{Roy03}
J.~S.~Kong and V.~P.~Roychowdhury, e-print arXiv:0711.3263v2.
 
\end{thebibliography}

\end{document}